\documentclass[11pt]{article}
\usepackage[margin=1in]{geometry}
\usepackage{lmodern}
\usepackage{sidecap}
\usepackage{xspace}
\usepackage{doi}
\usepackage{hyperref} 
\usepackage{tikz}
\usepackage{pgfgantt}
\usepackage{graphicx,pbox}
\usepackage{xcolor,hyperref,colortbl}
\usepackage{wrapfig}
\usepackage{enumitem}
\usepackage{gensymb}
\usepackage{comment}

\newcommand\pubnumber{ }
\newcommand\pubdate{\today}
\newcommand\pubblock{\rightline{\begin{tabular}{l} \pubnumber\\
          \pubdate \end{tabular}}}

\def\Title#1{\begin{center} {\Large #1 } \end{center}}
\def\Author#1{\begin{center}{ \sc #1} \end{center}}

\def\Address#1{\begin{center}{ \it #1} \end{center}}

\newcommand\snowmass{\begin{center}\rule[-0.2in]{\hsize}{0.01in}\\\rule{\hsize}{0.01in}\\
\vskip 0.1in Submitted to the  Proceedings of the US Community Study\\ 
on the Future of Particle Physics (Snowmass 2021)\\ 
\rule{\hsize}{0.01in}\\\rule[+0.2in]{\hsize}{0.01in} \end{center}}

\begin{document}
\begin{titlepage}
\snowmass
\pubblock

\Title{\textbf{RF Accelerator Technology R\&D} \\ \vspace{0.3cm} Report of AF7-rf Topical Group to Snowmass 2021}

\vspace{0.1cm}


\Author{\textbf{AF7-rf Conveners:}\\ Sergey Belomestnykh$^{1,2}$, Emilio A. Nanni$^{3,4}$,  and  Hans Weise$^{5}$} 




\Author{\textbf{Community Contributors:}\\ Sergey V. Baryshev$^{6}$, Pashupati Dhakal$^{14}$, Rongli~Geng $^{18}$, Bianca Giaccone$^{1}$, Chunguang Jing$^{7,8}$, Matthias Liepe$^{9}$, Xueying Lu$^{7,10}$, Tianhuan Luo$^{17}$, Ganapati~Myneni$^{11,12,13,14}$, Alireza~Nassiri$^{6,7}$, David~Neuffer$^{1}$, Cho-Kuen Ng$^{3}$, Sam~Posen$^{1}$, Sami Tantawi$^{3,4}$, Anne-Marie~Valente-Feliciano$^{14}$, Jean-Luc Vay$^{17}$, Brandon Weatherford$^{3}$, Akira~Yamamoto$^{15,16}$}

\Address{
$^{1}${Fermi National Accelerator Laboratory, Batavia, IL 60510, USA}\\
$^{2}${Stony Brook University, Stony Brook, NY 11794, USA} \\
$^{3}${SLAC National Accelerator Laboratory, Menlo Park, CA 94025, USA}\\
$^{4}${Stanford University, Stanford, CA 94305, USA}\\
$^{5}${Deutsches Elektronen-Synchrotron, Notkestrasse 85, 22607 Hamburg, Germany}\\
$^{6}${Michigan State University, East Lansing, MI, USA}\\
$^{7}${Argonne National Laboratory, Lemont, IL, USA}\\
$^{8}$ Euclid Techlabs LLC, 365 Remington Blvd., Bolingbrook, IL, USA\\
$^{9}$Cornell University, Ithaca, NY, USA\\
$^{10}$Northern Illinois University, DeKalb, IL, USA\\
$^{11}${BSCE Systems, Inc., Yorktown, VA, USA}\\
$^{12}${International Symposium On Hydrogen In Matter (ISOHIM), Yorktown, VA, USA}\\
$^{13}${Old Dominion University, Norfolk, VA, USA}\\
$^{14}${Jefferson Lab, Newport News, VA, USA}\\
$^{15}${KEK, Tsukuba, Japan}\\
$^{16}${CERN, Geneva, Switzerland}\\
$^{17}${Lawrence Berkeley National Laboratory, Berkeley, CA, USA}\\
$^{18}${Oak Ridge National Laboratory, Oak Ridge, TN, USA}\\
}



\end{titlepage}
\newpage

\tableofcontents
\newpage

\addcontentsline{toc}{section}{Executive Summary}
\section*{Executive Summary}

Accelerator radio frequency (RF) technology has been and remains critical for modern high energy physics (HEP) experiments based on particle accelerators. Tremendous progress in advancing this technology has been achieved over the past decade in several areas highlighted in this report. These achievements and new results expected from continued R\&D efforts could pave the way for upgrades of existing facilities, improvements to accelerators already under construction (e.g., PIP-II), well-developed proposals (e.g., ILC, CLIC), and/or enable concepts under development, such as FCC-ee, CEPC, C$^3$, HELEN, multi-MW Fermilab Proton Intensity Upgrade, future Muon Colloder, etc. Advances in RF technology have impact beyond HEP on accelerators built for nuclear physics, basic energy sciences, and other areas. Recent examples of such accelerators are European XFEL, LCLS-II and LCLS-II-HE, SHINE, SNS, ESS, FRIB, and EIC. To support and enable new accelerator-based applications and even make some of them feasible, we must continue addressing their challenges via a comprehensive RF R\&D program that would advance the existing RF technologies and explore the nascent ones.

The ongoing RF technology R\&D efforts closely follow the decadal roadmap that was developed in the framework of the DOE General Accelerator R\&D (GARD) program in 2017~\cite{GARD-RF-Strategy}. The roadmap reflects the most promising research directions that can potentially enable future experimental HEP programs. Similar to the DOE GARD roadmap, the European particle physics community developed a roadmap for European accelerator R\&D presented in a report, which includes a roadmap for high-gradient RF structures and systems~\cite{ESPP-RF-roadmap}. The two roadmaps cover similar RF technology topics thus presenting opportunities for collaboration between the U.S. and European institutions.

Here we highlight recent progress in selected key RF technology areas.

The main breakthroughs in the \textbf{Superconducting RF (SRF) cavity performance} were in the area of developing advanced surface treatments. Following the discovery about 10 years ago, nitrogen doping was further developed and quickly adopted for SRF cavities of LCLS-II. This success triggered world-wide collaboration on studying different surface treatments and resulted in several new treatment recipes, including variants of nitrogen doping for LCLS-II-HE and PIP-II, mid-temperature treatment, nitrogen infusion, and 2-step low-temperature baking. Further experimental and theoretical research is needed to fully understand physics of RF superconductivity in the thin surface layer. This understanding would allow precise tuning of treatment recipes to specific accelerator applications. However, in the medium accelerating gradient range (15 to 25~MV/m), already developed recipes have an impact on the performance of prototype SRF cavities for PIP-II and CEPC and further developments will impact FCC-ee, multi-MW Fermilab Proton Intensity Upgrade, and possibly a muon collider. For machines that are proposed to operate at higher gradients, new recipes, e.g., 2-step baking, will push the cavity performance to 55--60~MV/m in standing wave structures and to $\sim70$~MV/m and beyond in traveling wave SRF structures. This would enable future upgrades of the ILC or a new, more compact, SRF linear collider HELEN. In combination with other cost saving measures under development -- medium grain niobium material, novel high-efficiency RF sources, advanced resonance control, etc. -- the proposed SRF R\&D program would make future HEP facilities more affordable.

On a longer R\&D time scale, it is very important to continue studies of advanced thin film technologies and innovative materials for SRF cavity applications. The most advanced of such materials is Nb$_3$Sn. The superheating field of Nb$_3$Sn corresponds to a maximum accelerating gradient of approximately 100~MV/m (for standing wave structures), well above that of niobium. Even with very limited investments in this material so far, the best Nb$_3$Sn single-cell cavities can already reach $\sim25$~MV/m. Also, having a critical temperature about twice of the niobium critical temperature, Nb$_3$Sn can operate at temperatures $\sim4.5$~K while maintaining high quality factors, thus offering significant capital and operational savings in cryogenic systems. Research to further exploit the material and reach its full potential should continue, including novel surface treatment and deposition techniques.  Beyond Nb and Nb$_3$Sn, several other superconducting materials and advanced thin film technologies are being investigated.

The most notable development in the \textbf{high-gradient normal conducting RF} area is novel parallel-coupled C-band structures operating at liquid nitrogen temperature for the recently proposed C$^3$ collider and other applications. The highly optimized cell shape of the standing wave structure and increased electrical conductivity of copper at 80~K result in significantly improved shunt impedance and hence reduced RF power. The lower thermal stresses in the material and improved material strength reduce probability of breakdown, allowing to design the collider with an accelerating gradient of 120~MV/m. Prototype one-meter structures have been fabricated and tested at high gradient and at cryogenic temperatures. The next step is to develop an HOM damped and detuned structure design to mitigate the effect of the long-range wakefields in C$^3$. 

Optimization of cavity design, temperature, frequency and material properties has also allowed for the operation of structures with a gradient in excess of 200~MV/m, and cavities with a shunt impedances of a G$\Omega$/m. The powering of these structures with novel high-efficiency RF sources, which have also made significant progress, is opening new frontiers in beam brightness, beam manipulation, diagnostics, and acceleration. 

\textbf{Structure Wakefield Acceleration (SWFA)}, one of the advanced accelerator concepts, uses high-charge drive beam to excite intense wakefield in a structure. Then these wakefield accelerate a low-charge main beam. Use of short pulses of $\sim 10$~ns (more than an order of magnitude shorter than those typically used in pulsed high-power klystrons) in SWFA structures could mitigate RF breakdown and result in higher gradients, exceeding 150~MV/m and perhaps reaching 300~MV/m. Much progress has been made in recent years in developing dielectric structures (including dielectric tubes, dielectric slabs and dielectric disk-loaded structures) and metallic periodic structures in both X-band and millimeter-wave band. Structures with novel topologies, such as metamaterial structures, photonic bandgap structures, and photonic topological crystals, have been tested successfully.

\textbf{RF systems and sources}, a significant portion of the facility infrastructure, power and control the beams that are delivered to HEP experiments. Dedicated efforts to improve peak power, average power, and efficiency have shown remarkable progress on all fronts with the adoption of new beam bunching methods, new manufacturing techniques, and multi-beam devices. New concepts for RF components from cavity tuners to windows to pulse compressors continue to improve the power levels and efficiency with which we deliver RF power to accelerators. Importantly, the overall RF system performance does not only rely on power and efficiency, but also on precision and stability. Modern controls and low level RF (LLRF) electronics have made tremendous progress with RF phase and amplitude stability, fast feedback, and beam positioning. Combined with real time AI/ML controls significant improvements in beam brightness and luminosity are within reach.

Discussion about \textbf{Innovative Design and Modeling} highlighted aspects of high current and high brightness sources, bright beams and wakefields, accelerator modeling, and some cavity R\&D issues. The exchange between a number of key experts triggered the writing of several Snowmass White Papers.  
A Beam and Accelerator Modeling Interest Group (BAMIG) was formed, consisting of 25 key players from 13 U.S. laboratories and Universities. The common Snowmass White Paper emphasizes the importance of computational tools for the critical design, commissioning, operation, and upgrading of accelerator facilities.

Most advanced and often sophisticated high-performance computing tools are required to support R\&D activities.  Efforts in code writing are often local and somewhat uncoordinated which leads to duplication, to non-exiting interoperability, to challenges with respect to sustainability, and to the simultaneous retirement of codes and code owners. The need for advanced simulation studies, the long-term support for code development and maintenance, strengthening of collaborative efforts among laboratories and universities, the enabling of ‘virtual prototyping’ of accelerator components, the improvement of real-time simulations, -- all this is recognized as vital for new accelerator development.  In the White Paper modeling needs are summarized according to the fields of RF-based acceleration, plasma- and structure-based wakefield acceleration, petavolts per meter plasmonics and plasmonic acceleration, materials modeling for accelerator design, structured plasmas, and superconducting magnets. The author team describes each field and lists important references. Sustainability, reliability, user support and training are addressed. The path towards a community ecosystem is sketched. A Center for Accelerator and Beam Physics Modeling is proposed, and the envisaged activities are listed. 

While the GARD roadmap continues to serve as a community-developed guidance for the RF technology R\&D, it would benefit from some mid-course corrections. Based on the discussions and submitted White Papers, we present the following key directions that should be pursued during the next decade:
\begin{itemize}
    \item Studies to push performance of niobium and improve our understanding of SRF losses and ultimate quench fields via experimental and theoretical investigations;
    \item Developing methods for nano-engineering the niobium surface layer and tailoring SRF cavity performance to a specific application, e.g., a linear collider, a circular collider, or a high-intensity proton linac;
    \item Investigations of new SRF materials beyond niobium via advanced deposition techniques and bringing these materials to practical applications;
    \item Developing advanced SRF cavity geometries to push accelerating gradients of bulk niobium cavities to $\sim70$~MV/m for either upgrade of the ILC or compact SRF linear collider;
    \item Pursuing R\&D on companion RF technologies to mitigate field emission, provide precise resonance control, enable robust low level RF systems for high gradient and high Q accelerators, etc.;
    \item Research on application of SRF technology to dark sector searches;
    \item R\&D on high-gradient normal conducting RF structures operating at cryogenic temperatures with a gradient of $>150$~MV/m as a promising way toward a compact linear collider;
    \item R\&D on high frequency and multi-frequency structures to transcend limits on shunt impedance and accelerating gradients;
    \item Development of higher gradient normal conducting cavities in exotic environments (e.g. strong magnetic fields);
    \item Investigation of novel materials and manufacturing techniques to improve high gradient performance and remove design constraints;
    \item Developing high efficiency, low-cost RF sources that would benefit many operating and practically every future intensity or energy frontier machine;
    \item Studies dedicated to industrialization and cost reduction of fabricating RF components and systems;
    \item Continue research on advanced SWFA structures to bring them closer to practical applications through higher gradients ($>$300 MV/m), higher repetition rates and with damping features;
    \item Experimental and theoretical research to further our understanding of the RF breakdown physics;
    \item Continued development of computational tools for multi-physics and virtual prototyping;
    \item Developing a community ecosystem for accelerator and beam physics modeling that would incorporate comprehensive set of high-performance simulation tools for RF-based accelerators.
\end{itemize}

To support these key research directions, there is a need to upgrade and add new capabilities to the existing R\&D and test facilities to investigate the new concepts and help integrate them into systems with ready access to researchers.  Collaborative efforts at National Laboratories and universities have provided a broad spectrum of sources and manufacturing facilities that has enabled this progress. However, much of this infrastructure is aging and in need of rejuvenation. Without adequate investment in the facilities, further progress in advancing RF technologies will be hindered.

The workforce that supports the existing capabilities and facilities is currently insufficient. A significant portion of this workforce is approaching the end of their career. Bringing the next generation of staff into these facilities is a struggle. Additional resources and a strategy are urgently needed for education, training, and knowledge transfer.  

\newpage

\section{Introduction} 

Radio frequency (RF) technology is at the cornerstone of particle acceleration. With a few exceptions, all modern accelerators use resonant RF structures to impart energy onto (accelerate) charged particles of different species, from electrons to heavy ions. Such structures operate at resonant frequencies from tens of megahertz to tens of gigahertz. In addition to acceleration, RF structures are utilized for deflecting particles or ``crabbing'' them to mitigate luminosity loss in colliding schemes with crossing angles.

RF technology continues to address challenges of new accelerators, in particular those proposed for particle physics experiments. Tremendous progress in advancing this technology has been made over the past decade in several areas highlighted in this report. The achievements and new results expected from continued R\&D efforts could pave the way for upgrades of existing facilities, improvements to accelerators already under construction (e.g., PIP-II), well-developed proposals (e.g., ILC, CLIC), and/or enable concepts under development, such as FCC-ee, CEPC, C$^3$, HELEN, multi-MW Fermilab Proton Intensity Upgrade, future Muon Collider \cite{MCF}, etc. Advances in RF technology have impact beyond HEP on accelerators built for nuclear physics, basic energy sciences, and other areas. Recent examples of such accelerators are the European XFEL, LCLS-II and LCLS-II-HE, SHINE, SNS, ESS, FRIB, and EIC. To support and enable new accelerator-based applications and even make some of them feasible, we must continue addressing their challenges via a comprehensive RF R\&D program that would advance the existing RF technologies and explore the nascent ones.

In 2017, a decadal roadmap for RF technology R\&D was developed in the framework of the DOE General Accelerator R\&D (GARD) program~\cite{GARD-RF-Strategy}. The roadmap was put together by a team of leading researchers in the field from national labs and universities, both domestic and international. It provided a community-directed guidance and reflects the most promising research directions for advances that enable future experimental high energy physics programs. 

Recently, the 2020 update of the European Strategy for Particle Physics~\cite{ESPP2020} emphasized the importance of accelerator R\&D. The followed up effort on developing a roadmap for European accelerator R\&D culminated in a report, which includes a roadmap for high-gradient RF structures and systems~\cite{ESPP-RF-roadmap}. The European roadmap covers topics similar to the GARD roadmap and hence there is a lot of synergy and opportunities for collaboration between the U.S. and European laboratories and universities.

The GARD RF technology roadmap guided discussions of the Snowmass AF7-rf topical group, as most of its content remains timely and relevant. In this report we summarize the activities of the topical group. Further, based on discussions, presentations during miniWorkshops and seminars, and submitted contributed papers, we lay out a framework for key RF technology R\&D directions, which are largely aligned with, but in some areas extend beyond the GARD roadmap, and reflect the technological advances needed to upgrade existing or create new facilities for High Energy Physics.

\section{Topics}

Based on submitted Letters Of Interest (LOIs), the AF7-rf topical group selected three main topics for in-depth discussions. Accordingly, the group hosted three miniWorkshops on ``RF Systems and Sources,'' ``Innovative Design and Modeling,'' and ``Cavity Performance Frontier.'' The goal of these workshops was to discuss the submitted LOIs, learn about future research plans, discuss important HEP technical challenges, and strategize as a community for our collective contributions to Snowmass. In this section we discuss the outcomes from the miniWorkshops and review the submitted White Papers in the three topics.

\subsection{Cavity Performance Frontier}  

The miniWorkshop on Cavity Performance Frontier~\cite{CavityPerformanceWS} focused on different types of accelerating structures for future HEP machines. Three distinct technologies were considered: Superconducting RF (SRF) cavities; High-gradient normal conducting RF structures; Structure Wakefield Acceleration (SWFA) technology. The community discussions during this miniWorkshop and subsequent meetings resulted in several Snowmass White Papers on these RF technologies.

\subsubsection{Superconducting RF cavities}

Superconducting radio frequency (SRF) technology is critical for several frontiers of experimental high energy physics. SRF cavities make up the vast majority of the PIP-II linac, which will provide high-intensity proton beam to drive the neutrino facility LBNF/DUNE \cite{LBNF-DUNE}. One of the options for the multi-MW Fermilab Proton Intensity Upgrade is based on an SRF linac~\cite{8GeVSRFlinac}. SRF systems accelerate beams of the LHC and will provide crabbing at the interaction regions to boost luminosity of the HL-LHC \cite{Apollinari:2017lan}. SRF structures would provide energy for beams of the proposed Higgs factories, including ILC~\cite{ILC_TDR-v3-I,ILC_TDR-v3-II,ILC_Snowmass2021}, FCC-ee~\cite{FCC:2018byv}, CEPC~\cite{CEPC_CDR}, and HELEN~\cite{HELEN}. In addition, SRF cavities are being explored not only for particle beam acceleration but for detection in the next generation of dark sector searches~\cite{braine2020,Tobar_LOI,DarkSRF,HarnikSearchesSRF-Snowmass2021}, as well as for quantum computing, which could be extremely beneficial for HEP applications~\cite{Harnik_LOI,Matchev_LOI}. To sum up: the SRF technology is indispensable for the future of high energy physics and will continue enabling new HEP experiments. Therefore, research and development in this field is crucial. Continued improvements in cavity performance make new scientific applications feasible when they would have otherwise been either unachievable or too expensive. 

The White Paper~\cite{SRF-key-directions} outlines the key future research directions, which largely continue to align with the GARD roadmap. These include: 

\begin{itemize}
    \item Studies pushing the performance of niobium, including doping, multi-step heat treatment, flux expulsion and flux losses. 
    \item Developing methods for nano-engineering the niobium surface layer and tailoring it for specific applications.
    \item Furthering our understanding of RF losses and ultimate quench fields of niobium via experimental and theoretical investigations of the physics of RF surface resistance and penetration of flux into superconductors at high fields.
    \item Research of new SRF materials beyond niobium -- such as Nb$_3$Sn (through various preparation techniques), layered structures, and other promising superconductors -- via advanced deposition techniques and bringing these materials to practical applications.
    \item Developing advanced cavity geometries to push accelerating gradients of bulk niobium cavities to $\geq70$~MV/m and pursuing R\&D on companion RF technologies to mitigate field emission, provide precise resonance control, etc.
    \item Investigating application of SRF technology to dark sector searches.
\end{itemize}
These and some other directions were presented at the miniWorkshop, and several White Papers were submitted following the presentations and discussions. Here we briefly summarize the White Papers.

The main recent breakthroughs were in the area of advanced surface treatments of cavities made from bulk niobium. After the discovery of nitrogen doping about 10 years ago, this method was further developed and quickly adopted for SRF cavities of the LCLS-II linac. This success triggered a world-wide collaboration effort on studying different surface treatments and resulted in several new recipes, including variants of nitrogen doping for LCLS-II-HE and PIP-II, mid-temperature treatment, nitrogen infusion, and 2-step low-temperature baking. Further experimental and theoretical research is needed to fully understand the physics of RF superconductivity in a thin surface layer. This understanding would allow precise tuning of treatment recipes to specific accelerator applications. However, in the medium accelerating gradient range (from 15 to 25~MV/m), already developed recipes have an impact on the performance of prototype SRF cavities for PIP-II and CEPC~\cite{CEPC_CDR}. Further developments will impact FCC-ee~\cite{FCC:2018byv}, multi-MW Fermilab Proton Intensity Upgrade~\cite{8GeVSRFlinac}, and possibly a muon collider. For machines that are proposed to operate at higher gradients, new recipes, e.g., 2-step baking, will push the cavity performance to 55--60~MV/m in standing wave structures and to $\sim70$~MV/m and beyond in traveling wave SRF structures~\cite{TW_optimization}. This would enable future upgrades of the ILC~\cite{ILC_Snowmass2021} or building a new, more compact, SRF linear collider HELEN~\cite{HELEN}. In combination with other cost saving measures under development -- medium grain niobium material~\cite{Mid-grain_Nb_Snowmass}, novel high-efficiency RF sources, advanced resonance control, etc. -- the proposed SRF R\&D program would make future HEP facilities more affordable.

Field emission (FE) is one of the limiting factors for SRF cavities operating in accelerators. No full-scale SRF cavity has ever been operated at a gradient of $>$~40~MV/m. Hence, developing new cleaning methods \cite{GengSnowmass},  better cryomodule integration techniques, e.g., by employing robot-assisted assembly, and post-integration mitigation methods remain important R\&D topics. White Paper~\cite{MartinelloSnowmass} describes recent developments and future perspectives for one of the FE mitigation methods, in situ plasma processing. Plasma processing of SRF cavities was developed originally at Oak Ridge National Laboratory (ORNL), where it was demonstrated that a plasma of mixture of neon and a small percentage of oxygen reduces hydrocarbon-related field emission in the Spallation Neutron Source (SNS) cavities. Starting from this successful experience, plasma processing studies are being conducted at several laboratories for different accelerating structures. It whas been demonstrated that plasma cleaning can completely eliminate multipacting in TESLA-type cavities, therefore potentially saving all the effort usually needed to process multipacting in these structures. However, to gain the maximum benefit from the processing, different and more aggressive gas mixtures should be explored for cleaning the inner cavity surface from a large variety of contaminants. An aggressive R\&D effort should then be pursued with the goal of extending the applicability of plasma cleaning to cavities and cryomodules of different types, increasing the processing effectiveness against a variety of field emitters, and optimizing the processing effectiveness against multipacting.

Recent technological advances pushed the performance of best SRF cavities close to the DC superheating field of $\approx240$~mT and dramatically boosted the quality factors in the medium field range. These results pose a question whether the best Nb cavities are already close to the fundamental limit and reaching the accelerating gradients $\sim100$~MV/m at 2 K may require new SRF materials like Nb$_3$Sn. Currently, this question cannot be answered since the theoretical SRF performance limits at GHz frequencies are unknown. The estimates are based on extrapolations of classical old results to the parameter space where those theories are not applicable. The authors of the White Paper~\cite{GurevichSnowmass} suggest a theoretical SRF research program based on modern theories of non-equilibrium superconductivity under strong electromagnetic field which have been developed in the last 20 years with the following key directions:
\begin{itemize}
    \item Establishing the $Q$ limit: mechanisms of nonlinear surface resistance and the residual resistance in a non-equilibrium superconductor under a strong RF field.
    \item Establishing the SRF breakdown field limit: Dynamic superheating field and its dependencies on frequency, temperature and concentration of impurities.
    \item RF losses due to trapped vortices and extreme dynamics of ultrafast vortices driven by strong RF Meissner currents in SRF cavities.
    \item Optimization of SRF performance due to surface nano-structuring of the cavity surface: superconductor-insulator-superconductor (SIS) multilayers and impurity management.
\end{itemize}

These theoretical activities will be performed in collaboration with SRF experimental groups to meet the needs of the HEP community. Solving the outlined above challenging problems would not only establish the much-needed fundamental SRF limits but also help guide the materials optimization to boost the cavity performance and extend the high energy physics research to other areas, particularly the emerging quantum information technologies involving high-$Q$ SRF cavities.

Three White Papers discuss new SRF materials beyond niobium~\cite{ValenteSnowmass,PosenSnowmass,BarziSnowmass}. Research on advancing performance of superconducting cavities beyond the capabilities of bulk niobium follows the three thrusts focused on developing the next-generation thin-film based cavities. The first line of developments aims at investigating Nb/Cu coated cavities that perform as good as or better than bulk niobium at reduced cost and with better thermal stability. Recent results with greatly improved accelerating field and dramatically reduced $Q$-slope show the potential of this technology for many applications. While the performance of such cavities still fall short with respect to bulk niobium at high gradients, high current storage ring colliders such as FCC, EIC and CEPC, where the frequency is typically lower and the gradients are modest, could benefit from the cost savings and operational advantages of this technology.

The second research thrust is to develop cavities coated with materials that can operate at higher temperatures and/or sustain higher fields. A proof-of-principle has been established using Nb$_3$Sn~\cite{PosenSnowmass}. The superheating field of Nb$_3$Sn corresponds to a maximum accelerating gradient of approximately 100~MV/m (for standing wave structures), well above that of niobium. Even with very limited investments in this material so far, the best Nb$_3$Sn cavities can already reach $\sim25$~MV/m. Also, having a critical temperature about twice of the niobium critical temperature, Nb$_3$Sn can operate at temperatures $\sim4.5$~K while maintaining high quality factors, thus offering significant capital and operational savings on cryogenic systems. Research is now needed to further exploit the material and reach its full potential, perhaps with novel surface treatment and deposition techniques. Beyond Nb$_3$Sn, several other superconducting materials are being investigated for SRF cavity applications.

The third line of research is to push SRF performance beyond the capabilities of the superconductors alone by employing multi-layer coatings. In parallel, developments are needed to provide quality substrates, innovative cooling schemes and cryomodule design tailored
to SRF thin film cavities.

\subsubsection{High-gradient normal conducting RF structures}
\label{sec:High-gradient-RF}

High-gradient normal conducting RF technology for linear colliders has been under development since the 1970's. Due to high RF power dissipation in the cavity walls, the accelerating structures must operate in pulsed mode with short pulses of $< 1 \mu$s. Also, as the dissipation per unit length of the structure scales as $1/\sqrt{\omega}$, high frequencies are preferable for such structures. In this section, we review structures for two normal-conducting linear colliders -- CLIC~\cite{CLIC_Snowmass} and C$^3$~\cite{nanni2021c,C3_demostrator} -- that have been presented and discussed during different Snowmass events and described in White Papers. As usable gradients in these structures are limited by the rate of RF breakdowns, understanding the RF breakdown physics is very important. White Paper~\cite{Norem_Snowmass} identifies the dominant mechanisms of vacuum arcs, critical issues and desirable aspects of an R\&D program to produce a more precise and general model.

CLIC is based on a novel two-beam acceleration (TBA) scheme. The main linacs uses 12 GHz travelling wave accelerating structures with a tapered inner aperture diameter to accelerate particles from 9 GeV to the collision energy. The copper structures operate in the range of 70 MV/m to 100 MV/m. RF power for the main linacs is provided by a high-current, low-energy drive beam -- that runs parallel to the colliding beam -- through a sequence of power extraction and transfer structures (PETS). The RF power generated in the PETS is then transferred to the accelerating structures using a waveguide network. This TBA concept significantly reduces the cost and power consumption compared with powering the structures directly by klystrons~\cite{CLIC_Snowmass}. 

The performance of the CLIC structures have been validated in a series of dedicated tests that included an experiment at CTF3 at CERN to determine the effect of heavy beam loading. Fully assembled two-beam modules have been tested with and without beam. The CLIC RF design parameters are well understood and were reliably reproduced in tests. Further studies will focus on optimizing cost and energy efficiency of the RF system.

The most notable recent development in this field is novel parallel-coupled C-band structures operating at liquid nitrogen temperature for the proposed C$^3$ collider~\cite{nanni2021c} and other applications. The highly optimized cell shape of the standing wave structure and increased electrical conductivity of copper at 80~K result in significantly improved shunt impedance and hence reduced RF power. The lower thermal stresses in the material and improved material strength reduce probability of breakdown, allowing to design the collider with an accelerating gradient of 120~MV/m. Prototype one-meter structures have been fabricated and tested at high gradient and at cryogenic temperatures. The next step is to develop an higher-order-mode damped and detuned structure design to mitigate the effect of the long-range wakefields in C$^3$. Detuning of higher order modes will be achieved by modifying the geometry of each cavity while maintaining constant frequency of the fundamental mode. For damping, longitudinal damping slots in quadrature will be added to the structure design. A C$^3$ demonstration R\&D plan is discussed in White Paper~\cite{C3_demostrator}.

Optimization of cavity design, temperature, frequency and material properties has also allowed for the operation of structures with a gradient in excess of 200~MV/m, and cavities with a shunt impedance of a G$\Omega$/m. The powering of these structures with novel high-efficiency RF sources, which have also made significant progress, is opening new frontiers in beam brightness, beam manipulation, diagnostics, and acceleration.

Many of concepts that have been explored in the context of linear accelerators, could also benefit RF cavity operation in a muon cooling experiment. The needs and possibilities of interest are operation of normal conducting cavities in high magnetic fields, efficient power coupling, precise tuning, operation in poor vacuum, and cryogenic operation.
The muon collider (MC) RF cavity requirements align well with ongoing RF accelerator research, such as high gradient SRF cavities, high-efficiency RF power sources, etc. MC RF cavities also have their own unique features that require dedicated R\&D. One major topic is high gradient RF cavities in a multi-tesla superconducting solenoid field for muon ionization cooling. Two proof-of-principle cavities, one vacuum cavity with Be walls and the other a high-pressure gas-filled cavity, have demonstrated stable operation at 50~MV/m in a 3~T solenoid field. Still, significant work is needed to progress from these proof-of-principle cavities to the fully operating cavities for all of the cooling stages. One newly emerging scheme to overcome RF breakdown in strong magnetic fields is a copper cavity operated at cryogenic temperature. Investigation of these low temperature high-gradient structures is synergistic with other cryogenic NCRF R\&D (e.g., C$^3$). Another unique feature of the MC cavities is their operational environment, including the high radiation from muon decays and the strong fringe field due to the proximity of magnets. How SRF cavities perform in such an environment, and the corresponding mitigations if required, should be examined.

\subsubsection{Structure wakefield accelerators}
SWFA is one of the promising Advanced Accelerator Concept (AAC) schemes. As such, it is part of the 2016 DOE GARD AAC Roadmap and several other reports that include AAC research. However, as the basic SWFA structures are similar to conventional RF linac structures, this topic was included in the discussions under AF7-rf. Two White Papers have been submitted in SWFA area~\cite{Lu_SWFA_Snowmass,Jing_SWFA_Snowmass}. The first paper by Lu et al. is focused on Advanced SWFA structures R\&D. The second White Paper by Jing et al. describes the need of continuous and coordinated efforts for development of SWA for a potential energy frontier machine.

Like other AAC schemes, SWFA aims at raising the accelerating gradients beyond the limits of conventional RF accelerator technology. In an SWFA, a high-charge drive beam traverses a structure in vacuum and excites an intense wakefield. These wakefield can be used to accelerate a low-charge main beam. There are two SWFA schemes: i) collinear wakefield acceleration, or CWA, when the main beam follows the drive beam in the same structure and ii) two-beam acceleration, when the main beam is accelerated in a separate structure in parallel with the drive beam. Simply speaking, one can say that in SWFA structures an RF power source (e.g., klystron) is replaced with a short electron bunches of the drive beam. In conventional RF accelerators the usable accelerating gradient is limited to $\sim 150$~MV/m (even if the structures are cooled to cryogenic temperatures as in C$^3$~\cite{nanni2021c}) due to the rate of RF breakdown events. Use of short pulses of $\sim 10$~ns (more than an order of magnitude shorter than those typical for pulsed high-power klystrons) in SWFA structures could mitigate RF breakdown and result in higher gradients.

SWFA technology has been proposed for linear colliders and compact light sources. The most developed SWFA linear collider design is the Compact Linear Collider (CLIC) \cite{CLIC_Snowmass}. It is based on the TBA scheme. However, with a 244~ns pulse length, only about 100~MV/m loaded gradient can be used. Strictly speaking, the TBA scheme can be called SWFA only conditionally as the drive beam structure can be replaced with a klystron without altering the main accelerating structure, see more details in~\cite{CLIC_Snowmass}. We discussed the CLIC structure in section~\ref{sec:High-gradient-RF}. The Argonne Flexible Collider is based on the CWA scheme and proposed to operate with a shorter RF pulse of 20~ns to achieve a gradient of $\sim 300$~MV/m. Like conventional high-gradient RF structures, CWA accelerating modules could be used in compact X-ray FELs, e.g., \cite{Zholents_IPAC18}.

The following research directions are considered to be of importance for the next decade.

\textbf{Advanced wakefield structures:}  
Advanced structures with improved electromagnetic characteristics can dramatically improve SWFA performance. Much progress has been made in recent years in developing dielectric structures (including dielectric tubes, dielectric slabs, and dielectric disk-loaded structures) and metallic periodic structures in both X-band and in the millimeter wave band. Structures with novel topologies, such as metamaterial structures, photonic bandgap structures, and photonic topological crystals, have been tested successfully.

\textbf{Terahertz and sub-THz structures:} 
THz structures have the advantages of strong beam-structure interaction (high shunt impedance and thus high gradient) and small transverse size, which could lead to compact and cost-effective future colliders. When combined with bunch shaping techniques, THz structures could be ideal for high-gradient and high-efficiency wakefield acceleration. Recent advances in fabrication have made it possible to push SWFA into the mm-wave and THz regime.

\textbf{RF breakdown physics:} 
The physics of RF breakdown remains a hot topic for the high-gradient acceleration community. Previous studies were mostly carried out on conventional RF linac structures, with pulse lengths ranging from a few hundred nanoseconds to a few microseconds. Early SWFA experimental evidence has shown that short-pulse operation (a few ns) has the potential to dramatically increase the accelerating gradient. Further research, both experimental and theoretical, is necessary to bring insight into the RF breakdown physics in the parameter space relevant to SWFA.

\subsection{RF Systems and Sources} 

RF systems and sources are at the heart of almost any accelerator providing both the power and control to accelerate and manipulate charged particle beams. As the energy or intensity of these beams increase, the performance of the RF systems and sources plays an ever larger role in the performance of the overall accelerator complex. The importance of the RF systems and sources was recognized and key metrics were identified in the DOE GARD-RF Roadmap~\cite{GARD-RF-Strategy} for this decade. The Snowmass AF7-rf miniWorkshop on RF Systems and Sources \cite{RFSys1} was focused on high peak and high average power RF systems, RF components, and RF control. The presentations and discussion at the workshop covered both the present state of the art and future plans for the community. Several White Papers were submitted that directly cover the scope of these discussions: \cite{ives2022high,weatherford2022advanced,Scheinker22,kroc2022need,filippetto2022feedback,roser2022sustainability}. Additionally, facility White Papers for ILC~\cite{ILC_Snowmass2021}, CLIC~\cite{CLIC_Snowmass}, C$^3$~\cite{nanni2021c}, FCC~\cite{FCC} and the Muon Collider~\cite{Muon} identify key technologies to develop that could have significant impact on the performance or feasibility of these concepts.

\subsubsection{RF Systems}

The integrated RF system powering the accelerator plays a crucial role in the performance, stability and operational luminosity of any facility. The design of the low level RF controls, RF components and sources are all closely related and impact each others performance. While present generation facilities have sufficient precision and control, future HEP machines will be limited in luminosity or intensity without further advancement. 

\textbf{RF controls:}
Feedback and RF controls have made tremendous progress in recent decades with exceptional RF amplitude and phase stability, $\sim$50~nm position monitoring, and feedback latency on the order of 300~ns. However, pushing this to the next level will require significantly reduced feedback times to perform intra-train feedback~\cite{filippetto2022feedback}. This will be crucial for the development of advanced accelerator concepts that require more stringent tolerances. Interplay between digital and analog systems is critical for implementing these systems. In particular, for lowest possible latency all analog circuitry may be  required for the fast feedback loops. Interfacing these control systems with ML/AI enhanced operational software could significantly reduce machining tuning time, increase operational time and render operations more safe. \cite{scheinker2022adaptive,filippetto2022feedback}

\textbf{RF components:}
While RF components are ubiquitous in any accelerator, targeted improvements in certain systems would render significant advantages in performance and operational lifetime. Cavity tuners, RF power couplers, RF windows, pulse compressors, loads, feedthroughs, and HOM dampers are just some examples. Development of these components also offers a great opportunity to engage with industry and bring new products to the market (for example through SBIR programs). 

\subsubsection{RF Sources}

RF amplifiers for both CW and pulsed power systems are often thought of as “known quantities,” and while there is an abundance of research activity in accelerating structures, R$\&$D in high power RF sources is relatively uncommon. Certainly, industrial development of RF sources is required to improve the cost/capability challenge for RF power. However, doing so in isolation from ongoing high power RF research at universities and national laboratories, and the future plans for high energy physics facilities, is not possible. This is because there are not many commercially viable uses for megawatt-class RF amplifiers, and the devices that do exist are usually custom-designed for specific applications. It is unreasonable to expect that an industrial supplier will invest their own time and money to reduce cost in anticipation of a single-use system that may (or may not) be assembled decades in the future. Because of the long time frame, high technical risk, and undefined initial requirements associated with future facilities or future facility upgrades, any reasonable business plan would “price in” the impact of these uncertainties – so when a new facility is proposed, the RF power system will not be prohibitively expensive. Support for high-risk research in RF sources is desperately needed if we are to realize improvements in cost-capability that would make a dramatic difference. Such an effort must be led by national labs and academia because these institutions can tolerate the long term risk of this effort. Partnering with industry will also be critical to ensure that technology transfer is realized early on and address practical challenges to implementing new concepts.

An example of a successful lab-led R$\&$D initiative in high power RF sources has been the High Efficiency International Klystron Activity (HEIKA), established by CERN in 2014. This effort has supported worldwide collaboration to understand in detail how to optimize the electrical design of high power klystrons for maximum DC-to-RF efficiency. As a result of this work, new klystron design methods such as the Core Oscillation Method (COM) and Core Stabilization Method (CSM) were established. New designs for COM-based 8 MW and 50 MW X-band klystron prototypes have been completed by CERN and industrial partners~\cite{cai2021XBandKly}. In addition, a publicly available 2D klystron simulation code called KlyC was successfully developed, benchmarked, and made available to the RF source community~\cite{cai2019klyc}. The tools and design improvements arising from the HEIKA collaboration will be helpful in developing high efficiency (and therefore lower operational cost) prototypes for any future demonstration-scale collider, but more work still must be done to reduce the upfront construction costs of HPRF sources. The developments of HEIKA can also be considered in the context of a proposed facility. 

\begin{figure}[!ht]
    \centering
    \includegraphics[width=0.95\textwidth]{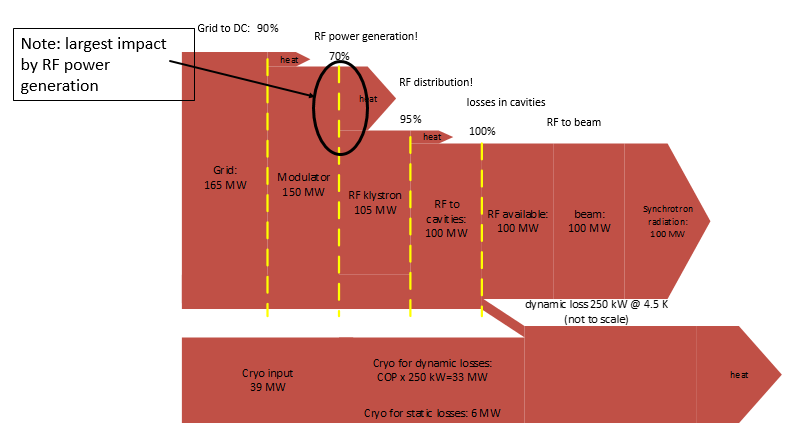}
    \caption{Energy conversion for the FCC-ee as presented and discussed at the AF7-rf workshop.\cite{WSJensen}}
    \label{fig:FCCpower}
\end{figure}

For example, if we consider the RF power needs of FCC-ee, Figure \ref{fig:FCCpower}, the single greatest inefficiency in the grid-to-beam power chain comes from the RF sources. Achieving the high average power and high efficiency for the RF sources will be critical in this case. Multi-beam klystrons for this purpose are in development with efficiencies that have slightly exceeded 70\% in prototypes. However, new concepts could push this significantly higher. Adding a stage to the multi-beam klystron where the beam is accelerated twice could increase efficiency to 85\%. This approach has the specific advantage of bunching the beam at low voltage (high perveance), allowing for a very compact RF bunching circuit, and has a bunched beam accelerated along the short DC voltage post-accelerating gap keeping the momentum spread small. Final power extraction occurs from a high voltage (low perveance) beam which boosts efficiency. There are also some additional advantages with the 2nd stage operated in DC, simplifying the modulator and increasing its conversion efficiency. Developing of such a concept would clearly impact the future prospects and performance of a primary candidate for a future high energy physics facility such as FCC-ee.

In the near term, research institutions should also identify and aggressively pursue new applications of RF sources and accelerator systems in the commercial, defense, and medical sectors; and to the degree possible, develop broadly useful RF source topologies that could have a need in high volume production. As one example, compact low-voltage klystron amplifiers are being developed for use in multiple linac-based radiography systems. If such a “building block” RF source can be optimized for use in small and large accelerator systems alike, and standardized as much as possible, then commercial opportunities and real competition between suppliers would drive significant cost reductions. Compare this to the approach of using custom-designed RF sources for a single facility, and the path to improved cost/capability is clear. 

Longer term, fundamental and exploratory research dedicated to RF sources and their components is essential for more than marginal improvements. The most reasonable approach involves optimizing the complete RF power chain, which naturally leads to using lower voltage modulators made from mass-produced commercially available components, which are simpler and require less infrastructure and maintenance.  Then, new RF sources are needed which can operate efficiently at low voltage and high current. Multiple-beam amplifiers leverage this concept, but this scaling approach can add complexity and does not really solve the fundamental problem - breaking the tradeoff between efficiency and perveance that is inherent in conventional linear-beam devices. Reconsidering RF sources in this way raises several interesting fundamental physics and engineering challenges. 

\textbf{High Average Power:}
Excellent progress is being made toward improving the performance of high average power RF sources. Multi-beam sources, novel device concepts, triodes, phase locked magnetrons, and distributed beam devices are all exploring innovated solutions for higher power, higher efficiency and lower cost~\cite{ives2022high,weatherford2022advanced}. Models of scaled production indicate that breaking through the GARD decadal goal of \$1/W average is within reach. Moving forward to prove these concepts will require increased production numbers and real-world deployment on accelerator systems to learn about lifetime and operational challenges. While industry has done a very good job pushing the power density of solid-state amplifiers that has made a 2 kW CW device available with an excellent power efficiency in 100's of MHz range, there is still room for R\&D on developing efficient power combiners (e.g., based on a combining RF cavity) in the 100's of kilowatt range.

\textbf{High Peak Power:}
The GARD decadal goal for high peak power RF sources is to achieve a large scale cost of \$2/kW peak. At the outset of the GARD Roadmap this RF source cost was one order of magnitude greater at \$20/kW peak. The adoption of new klystron designs with the core oscillation method, bunch align compress method and/or  better harmonic bunching control has significantly improved this outlook, with large scale studies indicating that the cost is in the \$7-10/kW peak range~\cite{CLIC_Snowmass}. Further, this will require the investigation of a host of possible paths from new electron sources for the beam, lower cost focusing magnets, lower cost manufacturing and distributed power generation. 

\subsection{Innovative Design and Modeling}

The Snowmass AF7-rf miniWorkshop on Innovative Design and Modeling \cite{Innov1, Innov2} highlighted aspects of high current and high brightness sources, bright beams and wakefields, accelerator modeling, and some cavity R\&D issues. The discussions and the exchange between workshop participants triggered the writing of several Snowmass White Papers \cite{Innov3,Innov4,Innov5,Innov6,Innov7,Innov8} .
\subsubsection{Accelerator Modeling Community White Paper}
A Beam and Accelerator Modeling Interest Group (BAMIG) was formed, consisting of over 80 subscribers to the mailing list, including 25 key players from 13 U.S. laboratories and Universities. The Snowmass White Paper~\cite{Innov9} emphasizes the importance of computational tools for the critical design, commissioning, operation, and upgrading of accelerator facilities. Most advanced and often sophisticated high-performance computing tools are required to support R\&D activities. Traditionally, large and complex simulations are based on codes which were often developed by a single accelerator physicist, and only some by interdisciplinary collaborations. The worldwide cooperation on research, design, construction and operation of largest accelerator facilities relies on small expert groups being in charge of modeling tools. Efforts are often local and somewhat uncoordinated which leads to duplication, to non-exiting interoperability, to challenges with respect to sustainability, and to the simultaneous retirement of codes and code owner. 

The BAMIG correctly claims that ever-increasing demands placed upon machine performance call for more complex simulation programs, a fact that will create more challenges unless the community decides to work more collaboratively and efficiently through the development of consortia and centers, also via collaboration with industry.

The White Paper refers to previous reports and recommendations over the last ten years. The need for advanced simulation studies, the long-term support for code development and maintenance, strengthening of collaborative efforts among laboratories and universities, the enabling of ‘virtual prototyping’ of accelerator components, the improvement of real-time simulations, - all this is recognized as vital for new accelerator development. Modeling needs are summarized according to the fields of RF-based acceleration, plasma- and structure-based wakefield acceleration, petavolts per meter plasmonics and plasmonic acceleration, materials modeling for accelerator design, structured plasmas, and superconducting magnets. The author team describes each field and lists important references. Ultraprecise, ultrafast virtual twins of particle accelerators are mentioned as the next frontier of the community. Interdisciplinary simulations address e.g. dark current induced problems, radiation levels and shielding, the modeling of positron production, particle/matter codes linked to CAD models, or – also including micro-physics models – the emission modeling of a high-brightness electron photocathode gun. End-to-end virtual accelerator modeling supports design but also operation, and ultimately, virtual twins of accelerators should be realized.

Cutting-edge accelerator technologies and related beam studies require strong development of software and algorithms. Machine learning and artificial intelligence make more and more substantial contributions. As a result, fast-executing, adaptive accelerator models become available. Physics-informed and -guided machine learning modeling is addressed.  Quantum computing can clearly add to the field. Start-to-end simulation of an accelerator using real beams with billions or more particles can potentially profit from quantum computers which support problem solving by focusing on the realization of selected algorithms with quantum circuits. 

The White Paper comments on computational needs, both hardware and software. Sustainability, reliability, user support and training are addressed. The path towards a community ecosystem is sketched. Workflows can help linking several codes or solving a problem with more physical processes, and benchmarking codes against each other or with known solutions.  I/O standardization and data repositories allowing to share results as well as making it available following Open Science practices can ease collaborative efforts. Centers and consortia are described as the path towards a coherent and consolidated effort in accelerator simulations. Examples in other disciplines are given. A Center for Accelerator and Beam Physics Modeling is proposed, and the envisaged activities are listed. 

BAMIG concludes as follows: Computer simulations will continue to be essential to particle accelerator research, design and operation. Its relative importance is even expected to grow, thanks to improvements in algorithms, computer hardware, and new opportunities in machine learning and quantum computing. These will enable accelerator modeling capabilities that include more physics that is integrated self-consistently to model accelerators with ever increasing fidelity and accuracy, toward the ultimate realization of the grand challenge of virtual twins of particle accelerators. A more collaborative and coordinated approach that enables the development of community ecosystems, adopting best practices in software developments and maintenance, is needed to meet the challenge in a realistic budget envelope and timeframe.

\subsubsection{Wakefield acceleration and related structures}
The miniWorkshop on Innovative Design and Modeling also covered the R\&D efforts towards wakefield acceleration and the respective continuous and coordinated efforts of Structure Wakefield Acceleration Development for an Energy Frontier Machine. The resulting White Papers are reviewed in Section 2.1.3 of the present report.

\subsection{Enable Facilities \& Upgrades}
\label{sec:Facilities}

The Implementation Task Force \cite{ITFreport} evaluated approximately twenty collider proposals that were submitted as part of this Snowmass process. Proposals spanned, electron, muon, gamma, and hadron colliders. The technologies included SRF and RF accelerators, wakefield accelerators and energy recovery concepts. Technical needs were also assessed for the initial proposals and upgrades. It is striking to see how essential RF accelerator technology development is to the realization of nearly every collider. The R\&D topics are generally covered and discussed in this report and those with specific impact on future colliders includes: RF sources, RF materials, RF structure designs for efficiency and gradient, RF components, RF power extraction, high brightness guns, high brightness sources, HOM dampers, and alignment/assembly of the accelerators. In many cases significant progress is needed to make these proposals a reality and the benefits are truly enourmous given the potential scale of these facilities. Beyond future colliders, accelerator RF technology enables such HEP applications as multi-MW proton drivers for neutrino experiments and RF-cavity-based dark sector searches.

\subsection{Synergies and Impact of HEP Accelerator R\&D on Other Fields} 

The great advantage of RF accelerator technology is that its impact outside of HEP is immediate and broad. RF accelerators are used for many different applications in science (e.g., nuclear and basic energy sciences), medicine, security, and industry. RF technology serves as the basis for light sources, electron and ion sources, X-ray sources used in scientific applications. The largest commercial sector for RF accelerators is the medical field where they are utilized primarily for X-ray radiation therapy or isotope production. RF accelerator technology improvements are presently driving a revolution in cancer therapy \cite{Boucher} that is enabling high dose rate X-ray, electron or proton therapy. Greatly increasing treatment options and potentially improving outcomes. These systems will require higher performance RF accelerator technology to be industrialized and could greatly advance the industrial readiness for a future major HEP facility based on RF accelerator technology.


\section{Acknowledgements}

The authors would like to acknowledge contributions of the RF community, especially those who participated in the discussions, submitted letters of interest and White Papers, and provided additional information for this report.

\addcontentsline{toc}{section}{Bibliography}

\bibliographystyle{atlasnote}
\bibliography{bibliography.bib}

\end{document}